\documentclass[epj]{svjour}
\usepackage{graphics,epsfig}
\begin{document}
\title{Bulk matter physics and its future at the Large Hadron Collider}
\author{B.~Hippolyte% etc
\thanks{e-mail: hippolyt@in2p3.fr}%
}                     % Do not remove
\institute{Institut Pluridisciplinaire Hubert Curien, D\'{e}partement de Recherches Subatomiques and Universit\'{e} Louis Pasteur, Strasbourg, France}
\date{Received: date / Revised version: date}
\abstract{
Measurements at low transverse momentum will be performed at the LHC for studying particle
production mechanisms in $pp$ and heavy-ion collisions. Some of the experimental capabilities for bulk matter
physics are presented, focusing on tracking elements and particle identification. In order to anticipate the
study of baryon production for both colliding systems at multi-TeV energies, measurements for identified
species and recent model extrapolations are discussed. Several mechanisms are expected to compete for
hadro-production in the low momentum region. For this reason, experimental observables that could be
used for investigating multi-parton interactions and help understanding the ``underlying event'' content in
the first $pp$ collisions at the LHC are also mentioned.
 % end of PACS codes
} %end of abstract
\maketitle
%
%
%
% SECTION 1: Introduction
%
\section{Introduction}
\label{intro_bohq08}
%
% General introduction for the topics
%
A significant physics programme is dedicated to large cross section observables at the LHC.
Starting from detector performances and the statistics for the first years of data taking, several items related to bulk matter are discussed in these proceedings.
We define here bulk matter physics as the physics dominated by soft processes occurring in heavy-ion and proton-proton collisions with a specific emphasis on multiple parton interactions.
We try to show how hadro-production at low transverse momentum ($p_{\rm{T}}$) can be used to investigate non-perturbative QCD phenomena such as quark recombination or, more generally, multi-parton dynamics.

%
% Specific introduction for the content
%
We start with presenting some experimental capabilities at the LHC in the second section:
several LHC experiments have tracking devices perfectly suited for identifying low to intermediate $p_{\rm{T}}$ (i.e. from $\sim$100~MeV/$c$  to  $\sim$6-8~GeV/$c$) particles and measuring production yields at mid-rapidity.
A few of these detector characteristics are summarized.
Due to its importance for the soft physics sector, information about the material budget in the central region for ALICE, ATLAS and CMS experiments is indicated as well.
The third section deals with baryon production, focusing on different aspects, such as thermal statistical description or baryon number transport.
In the fourth section, we concentrate on the hints of multi-parton dynamics, both in $p\bar{p}$ via underlying event studies at CDF as well as intermediate $p_{\rm{T}}$ measurements for baryons and mesons in $pp$ and A+A collisions at RHIC. Eventually, some model predictions for baryon over meson ratios vs $p_{\rm{T}}$ are shown at the nominal LHC energies.
%
%
%
% SECTION 2: Detector performances for soft physics at the LHC
%
\section{Detector performances for soft physics at the LHC}
\label{sec:det_per}
Detecting the bulk of particles requires good performances for tracking and identification at low $p_{\rm{T}}$. There is also a compromise to be found between a sufficient segmentation for dealing with high multiplicities and a low material budget for having good momentum resolutions. We propose here to check quickly key characteristics related to these issues for some of the LHC experiments. 
\subsection{Low $p_{\rm{T}}$ performances at mid-rapidity}
\label{sec:per_mid}
Most of the LHC experiments have a broad heavy-ion programme. However, each of them is not equally designed for facing the important multiplicities in Pb+Pb collisions at the nominal LHC energies.

%
% ATLAS
%
At mid-rapidity, the tracking device of the ATLAS experiment consists of a 3-layer pixel barrel followed by a Semi-Conductor Tracker (SCT) strip detector of 4$\times$2 layers ($|\eta|$~$<$~2.6), and complemented for e/$\pi$ separation by  a Transition Radiation Tracker (TRT) composed of 3 layers of straw-tubes interspersed with a radiator~\cite{:2008zz}.
Charged multiplicity and spectra measurements should benefit from the fine segmentation presented in Tab.~\ref{tab:atl_res}.

\begin{table}[!ht]
\begin{tabular}{lclclcl}
\hline\noalign{\smallskip}
subdetector  & radius range (cm) & position resolution ($\mu$m) \\
\noalign{\smallskip}\hline\noalign{\smallskip}
Pixel & 5 - 13 & 10 / 115 (r$\phi$/z)\\
SCT & 30 - 51 & 17 / 580 (r$\phi$/z)\\
TRT & 55 - 109 & 130 \\
\noalign{\smallskip}\hline
\end{tabular}
\caption{Radius ranges for the subdetectors of the ATLAS barrel tracker and corresponding position resolutions \cite{:2008zz}.}
\label{tab:atl_res}
\end{table}

Particle identification via linear energy loss ($dE/dx$) in the ATLAS central region would rely on a maximum of 11 clusters in the silicon layers. It is also important to note that the nominal magnetic field is of 2T which makes it challenging for low $p_{\rm{T}}$ particles to reach layers at large radius. Eventually, a high occupancy is expected in the TRT already for $pp$ collisions, possibly reaching 90\% for Pb+Pb ones (up to $dN_{\rm{ch}}/d \eta$=3000).\\

%
% CMS
%
Many studies were performed for anticipating the capabilities of the CMS experiment in the low $p_{\rm{T}}$ region and more specifically for heavy-ion collisions~\cite{Sikler:2007uh,Sikler:2008sv}.
They mainly involve the central tracker with a pixel detector of 3 layers, a Tracker Inner Barrel (TIB) of double-sided silicon strip sensors and a Tracker Outer Barrel (TOB) of silicon strip detectors in a 4T magnetic field.
More information on the detector layout can be found in Ref.~\cite{:2008zzk}.
It results in an excellent primary vertex and track impact parameter determination together with a maximum occupancy in the tracker of 30\% (layer 4) for Pb+Pb~\cite{D'Enterria:2007xr}.

\begin{figure}
  \epsfclipon
  \epsfig{figure=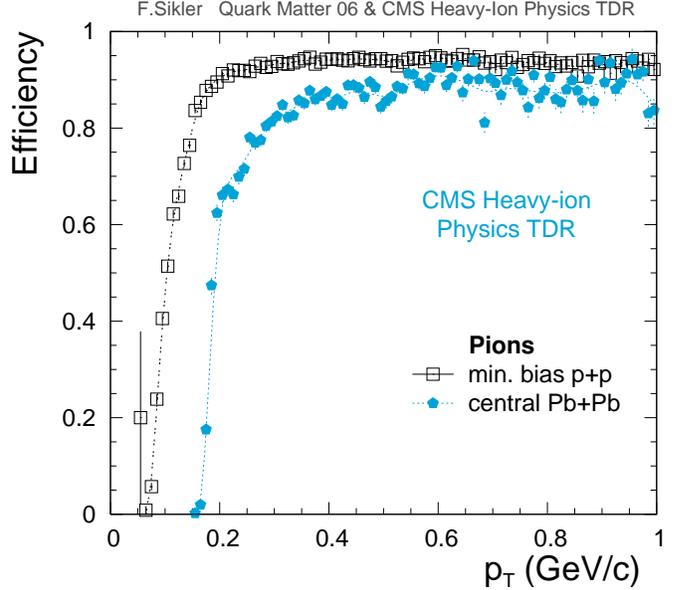, width=0.5\textwidth}
  \epsfclipoff
  \caption{Reconstruction efficiency as a function of $p_{\rm{T}}$ for pions using the CMS silicon tracker and within the pseudorapidity range $|\eta|<1$. Both efficiencies for minimum bias $pp$ events (open squares) and for central Pb+Pb collisions (full circles) are represented~\cite{Sikler:2007uh}.}
\label{fig:CmsEfficiencyPtPpPbPb}
\end{figure}

With 10 clusters (where 4 correspond to stereo measurements from the TIB) at mid-rapidity, it was demonstrated that particle identification (PID) via $dE/dx$ can be used to separate successfully charged pions, kaons and protons at low $p_{\rm{T}}$.
Strange neutral particles are also shown to be identified via invariant mass analyses. 
A modified algorithm for the pixel detector currently uses a helix parameterization instead of a straight line approximation.
Figure~\ref{fig:CmsEfficiencyPtPpPbPb} shows the reconstruction efficiencies as a function of $p_{\rm{T}}$ for a charged pion to leave 3 clusters in the pixel detector and being identified with the silicon strip barrels.
It contains both efficiencies for minimum bias $pp$ and central Pb+Pb collisions as presented in Ref.~\cite{Sikler:2007uh}.
Note that a multiplicity of $dN_{\rm{ch}}/dy|_{y=0}$=3200 was assumed for the central Pb+Pb simulations.
In the specific case of $pp$ collisions, the average reconstruction efficiencies reach 0.90, 0.90 and 0.86 respectively for pions, kaons and protons with $|\eta|<1.5$. The relative momentum resolution is quite significant at very low $p_{\rm{T}}$ ($\sim$10\% for protons at 0.2~GeV/$c$) but decreases to 6\% already at 1~GeV/$c$.\\

%
% ALICE
%
Tracking in the ALICE experiment is principally based on a large Time Projection Chamber (TPC) and an Inner Tracking System (ITS) using three different silicon technologies: pixel, drift and strip detectors (2 layers for each, respectively named SPD, SDD and SSD). 
There is a full azimuthal coverage and the pseudorapidity range for a charged particle is $|\eta|<0.9$~\cite{Aamodt:2008zz}.
The experiment was originally designed for central Pb+Pb multiplicities up to $dN_{\rm{ch}}/d \eta$=8000.
It has an excellent efficiency and resolution at low $p_{\rm{T}}$ benefiting from a moderate magnetic field of 0.5T.

\begin{figure}
  \epsfclipon
  \epsfig{figure=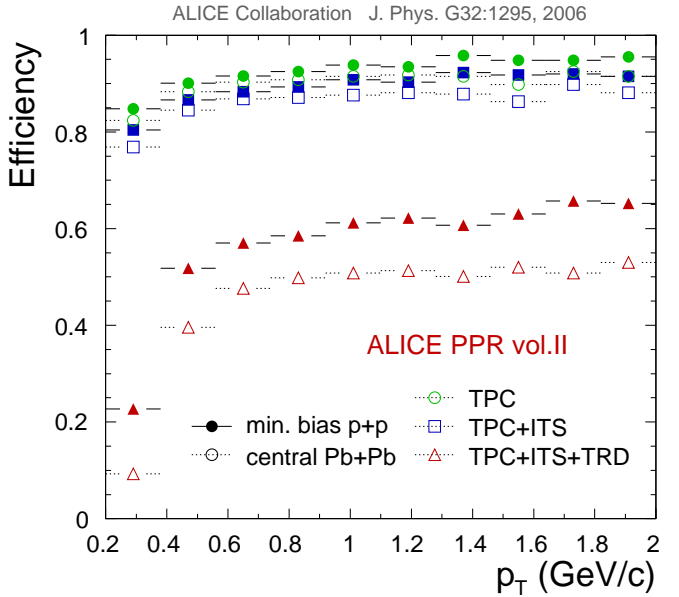, width=0.5\textwidth}
  \epsfclipoff
  \caption{Track-finding efficiency with different combinations of ALICE detectors at mid-rapidity for minimum bias $pp$ (full symbols) and central Pb+Pb (open symbols) collisions ($dN_{\rm{ch}}/d \eta$=6000)~\cite{Alessandro:2006yt}.}
\label{fig:AliceEfficiencyPt3Det}
\end{figure}

The TPC $dE/dx$ resolution is below 6\% and 8\% respectively for $pp$ and Pb+Pb collisions in the $p_{\rm{T}}$ range of [0.2-1.5]  GeV/$c$ with tracks having between 90 and 150 clusters.
Many other detectors will contribute to PID in a wide $p_{\rm{T}}$ range and not only for charged particles (for instance via Ring Imaging Cherenkov,  Time of flight determination or with a Transition-Radiation Detector for electron identification) but also for some neutral particles (e.g. via secondary vertex reconstruction and invariant mass analysis).
Detailed estimates for PID ranges and efficiencies can be found in Ref.\cite{Alessandro:2006yt}, such as Fig.~\ref{fig:AliceEfficiencyPt3Det} which illustrates the very good efficiency obtained for different combinations of ALICE detectors at mid-rapidity, for both minimum bias $pp$ and central Pb+Pb collisions.
%
%
% Material budget
%
\subsection{Material budget}
\label{sec:mat_bud}
\begin{figure}
\resizebox{0.5\textwidth}{!}{%
  \includegraphics{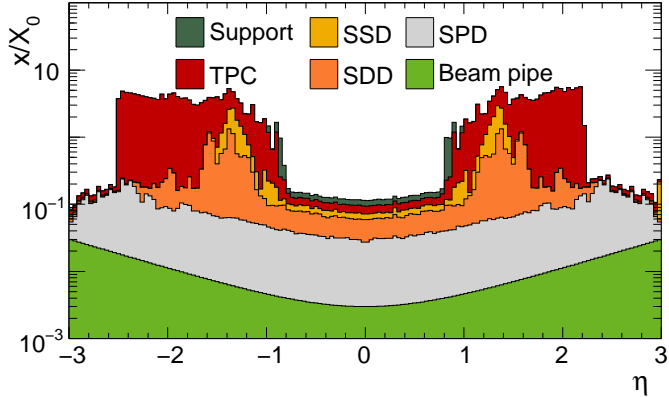}
}
\caption{Material distribution (in radiation length X$_{0}$) of the ALICE central tracking region, averaged over the azimuthal angle $\phi$. Contributions of individual subdetectors include service and support. Details of cumulative values at mid-rapidity and comparison with other LHC experiments can be found in Ref.~\cite{Revola:2008zz}.}
\label{fig:AliceRadiationLengthDetectorsLog}
\end{figure}

Cumulative material budget has a very important influence for low $p_{\rm{T}}$ measurements since it impacts on absorption and multiple scattering. A summary table, including ALICE, ATLAS and CMS information, is available in Ref.~\cite{Revola:2008zz} and indicates that ALICE tracker is the thinnest with 13\% of radiation length (i.e. approximately the third of the ATLAS or CMS one). In Fig.~\ref{fig:AliceRadiationLengthDetectorsLog}, we show the material distribution (in radiation length) of the ALICE central tracking region up to the TPC (outer radius of 2.6~m) and averaged over the azimuthal angle $\phi$.

A logarithmic scale is chosen for distinguishing properly the different contributions of the ITS subdetectors as well as the TPC in the central region.
Equivalent figures for ATLAS and CMS can be found in Refs.\cite{:2008zz,:2008zzk}.
%
%The reference for the radiation length distribution of ATLAS is figure 4.45 page 107 of the  JINST 3 (2008) S0800.
%
%
% SECTION 3: Baryon production mechanisms and studies
%
\section{Baryon production and asymmetry studies}
\label{sec:bar_mec}
Statistical thermal models reproduce quite successfully hadron yields over a large energy range. Therefore several extrapolations at LHC energies are already available for Pb+Pb collisions, anticipating a baryo-chemical potential very close to zero (see Ref.~\cite{Hippolyte:2006ra} and references therein). For $pp$ collisions, one can even try to distinguish between different correlated volumes within the canonical description~\cite{Kraus:2008fh}.

At the LHC, having a vanishing net-baryon density and a large gap between beam rapidities create ideal conditions for studying baryon number transport.
This can be done by measuring an asymmetry between baryon and anti-baryon production at mid-rapidity.
Predictions for such an asymmetry vary significantly, depending on models, when including or not baryonic string junctions~\cite{Montanet:1980te}. The experimental challenge is then to have sufficiently low systematical uncertainties for being able to draw conclusions~\cite{Christakoglou:2008an10}. It must be noted, however, that measurements consistent with no asymmetry were recently reported~\cite{Falkiewicz:2008dis}. 
%
% SECTION 4: Multi-parton dynamics
%
\section{Multi-parton dynamics in $pp$ and Pb+Pb}
\label{sec:mul_dyn}
In this section, we present briefly how multiple parton interactions were investigated by studying the underlying event structure in p+$\bar{\rm{p}}$ with the CDF experiment at the Tevatron and how this can be pursued with PID for $pp$ collisions  at the LHC.
Then basic features of the quark recombination are summarized with an illustration showing the possible interplay between coalescence and fragmentation in the intermediate $p_{\rm{T}}$ region.
\subsection{Underlying event studies}
\label{sec:bar_jun}
\begin{figure}
\resizebox{0.5\textwidth}{!}{%
  \includegraphics{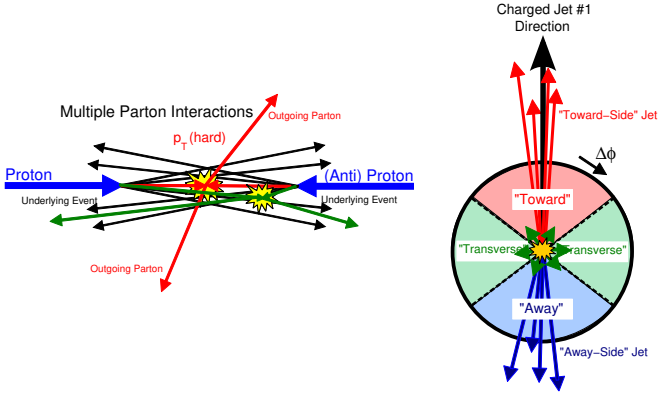}
}
\caption{Left: Illustration of the ``underlying event" modeled by PYTHIA for $p\bar{p}$ collisions by including multiple parton interactions (together with the hard 2-to-2 parton-parton scattering with a transverse momentum $p_{\rm{T}}$-hard, there is a second `semi-hard" 2-to-2 parton-parton scattering which contributes to particle production). 
Right: Illustration of the correlations in the azimuthal angle relative to the direction of the leading charged jet noted as ``Charged Jet \#1" in the event. The ``transverse" region, which is perpendicular to the axis of the hard 2-to-2 scattering, is very sensitive to the ``underlying event" component of the QCD Monte Carlo models \cite{Field:2006ek}.
}
\label{fig:CdfMpiPhiRegions}
\end{figure}

In $pp$ or $p\bar{p}$ at high energy, the emphasis is usually laid on the hard 2-to-2 parton-parton scattering with a high transverse momentum and fragmenting into jets of particles. However a significant fraction of particle production is coming from the ``underlying event", defined as the beam-remnants plus initial and final state gluon radiations as shown on the left side of Fig.\ref{fig:CdfMpiPhiRegions}.

Disentangling the different contributions can be complicated both phenomenologically and experimentally~\cite{Field:2006ek} but specific measurements from the CDF experiment gave insights on this issue~\cite{Affolder:2001xt}.
The idea is to distinguish several angular regions dominated by each contribution. Hence their definition is with respect to the back-to-back main jet direction as shown in the right side of Fig.\ref{fig:CdfMpiPhiRegions}:  the ``toward-side" region corresponds to leading (charged) jet direction with $|\Delta \phi| < 60^{\rm{o}}$ whereas the ``away-side" region is defined by $|\Delta \phi| > 120^{\rm{o}}$. Complementarily, the ``transverse" region is given by the interval $|\Delta \phi| \in [60^{\rm{o}} -120^{\rm{o}}] $ and is the most sensitive to the ``underlying event" and its soft structure.

Reproducing the main features of both contributions simultaneously is not an easy task for QCD Monte Carlo models since the underlying structure is simply not a minimum bias event. It triggered developments implying soft or semi-soft multiple parton interactions in order to reproduce the charged particle multiplicity or the transverse mean $p_{\rm{T}}$~\cite{subdetectortrand:2004ef}.
More constraints could be given using at the same time PID and a lower $p_{\rm{T}}$ cut at the LHC energies.
\subsection{Coalescence and hadro-production at RHIC}
\label{sec:coa_had}
\begin{figure}
\resizebox{0.5\textwidth}{!}{%
  \includegraphics{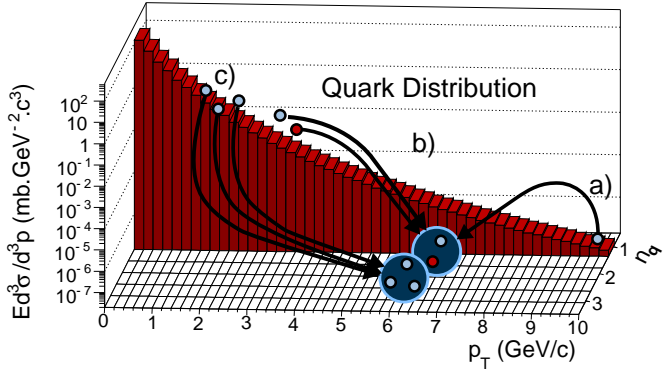}
}
\caption{Illustration of the possible competition between fragmentation and coalescence in the intermediate $p_{\rm{T}}$ range with: a) corresponds to a high $p_{\rm{T}}$ quark fragmentation into a meson, whereas b) and c) show respectively  the creation of a meson and a baryon via quark recombination.}
\label{fig:QuarkRecombination} 
\end{figure}

By going from $pp$ to A+A collisions at a sufficiently high energy in the center of mass, it is expected to switch from a dilute system of partons with high virtualities to a dense one with low virtualities eventually equilibrating to form a Quark Gluon Plasma.
If the phase space is filled with partons, hadronization via recombination/coalescence can be envisaged under ``the assumption of a thermalized parton phase"~\cite{Fries:2003kq}.
In the intermediate $p_{\rm{T}}$ region, in vacuo fragmentation competes with in medium recombination of lower momentum quarks as shown in Fig.~\ref{fig:QuarkRecombination}.
It illustrates the fact that due to a quickly decreasing quark momentum distribution, making a 6~GeV/$c$ pion from the recombination of two quarks at 3~GeV/$c$ is favoured with respect to the fragmentation of a 10~GeV/$c$ parton.
For the same reason, such a production mechanism was invoked to explain the high baryon over meson ratios in the transverse momentum range between 1 and 5 GeV/$c$ which were observed at RHIC as well as constituent quark scaling.
Developments including higher Fock states in order to take into account sea quarks and gluons and not only valence quarks can be found in Ref.~\cite{Muller:2005pv}.
%
%
% SECTION 5: Baryon / Meson ratios in $pp$ collisions at the LHC
%
\section{Baryon / Meson ratios in $pp$ collisions at the LHC}
\label{sec:rat_lhc}
\subsection{Reference for heavy-ion collisions}
\label{sec:ref_hic}
As reminded in section~\ref{sec:coa_had}, the baryon over meson (B/M) ratios measured at RHIC are considered as a first step for investigating recombination and coalescence mechanisms.
A summary of the measurements which were performed by the STAR experiment can be found in Ref.~\cite{Xu:hq08}.
Probing the differences between baryon and meson spectrum shapes in A+A at the LHC requires PID over a large $p_{\rm{T}}$ range. Calculations for the LHC are already available assuming a transverse radial flow extrapolation: the amplitude for mixed ratios are predicted to be comparable to the RHIC ones but the turnover should be shifted to slightly higher $p_{\rm{T}}$ values~\cite{Fries:2003fr}.
However, with the first LHC data being that of elementary $pp$ collisions, it also means that very detailed studies must be performed similarly for this colliding system as a reference~\cite{Hippolyte:2006ra}.
\subsection{Predictions from PYTHIA and EPOS}
\label{sec:pre_mod}
We report several extrapolations for $pp$ collisions at the LHC nominal energy of 14 TeV. Using PYTHIA v6.214, we start comparing the baryon production for a standard minimum bias extrapolation tuned by the ATLAS Collaboration and three underlying event descriptions. The descriptions only differ here by the $p_{T min}$ value used in PYTHIA and below which the perturbative parton-parton hard scattering cross section is assumed to vanish. The value is chosen be proportional to the square of the centre-of-mass energy to a chosen power $\epsilon$~\cite{subdetectortrand:1993yb:2003wg}.

\begin{figure}
\resizebox{0.5\textwidth}{!}{%
  \includegraphics{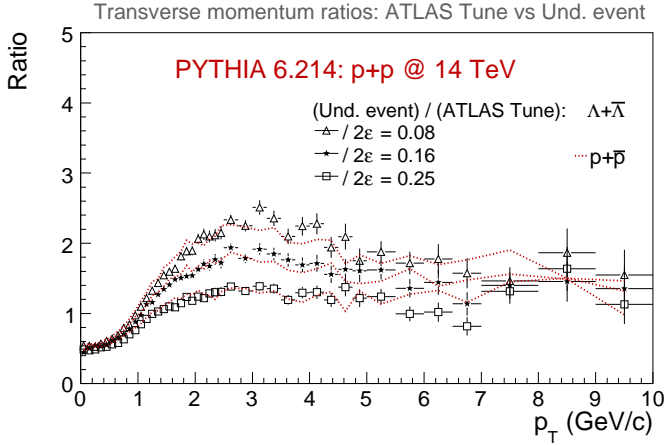}
}
\caption{Ratio of baryon productions in $pp$ collisions at 14~TeV in the centre of mass, for $\rm{\Lambda}$s (symbols) and protons (dashed lines) between PYTHIA ATLAS tune and underlying event descriptions: no $s$/$u$ flavour dependence is observed for these PYTHIA simulations and the underlying event contribution may be quite significant in the intermediate $p_{\rm{T}}$ region at LHC energies.}
\label{fig:Pythia6214LambdaPtRatioAtlasOverUe} 
\end{figure}

In Fig.~\ref{fig:Pythia6214LambdaPtRatioAtlasOverUe}, we show the results of these PYTHIA simulations in performing the ratio between the baryon production from the underlying event (``und. event") and minimum bias (``ATLAS tune") as a function of  $p_{\rm{T}}$. Both $p+\bar{p}$ and $\rm{\Lambda}+\bar{\rm{\Lambda}}$  are represented (respectively dashed lines and symbols).
With a decreasing $p_{T min}$ (corresponding to a decreasing power $2 \times \epsilon$), the possible contribution
of the underlying event increases exactly in the intermediate $p_{\rm{T}}$ region.
It indicates that the underlying event contribution can reach more than twice the one of minimum bias event at the top LHC energies. Note must be taken that no $s$/$u$ flavour dependence is observed, making the strange baryon interesting to study such an effect (topology identification of strange baryon decays is efficient in this $p_{\rm{T}}$ domain).

\begin{figure}
\resizebox{0.5\textwidth}{!}{%
  \includegraphics{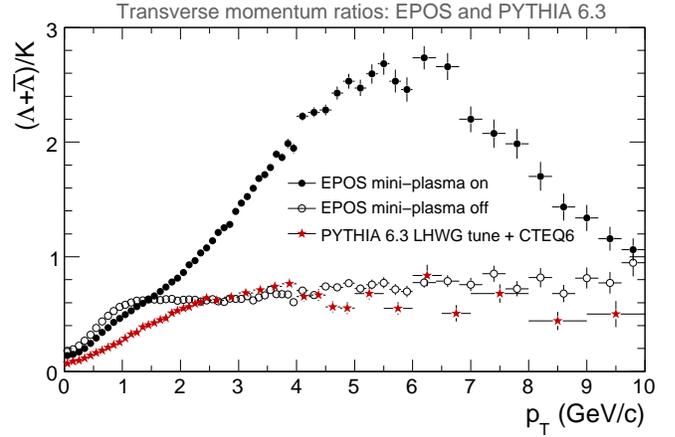}
}
\caption{Comparison of single strange baryon over meson ratio $(\rm{\Lambda}+\bar{\rm{\Lambda}})/\rm{K}$ at LHC top  energies for $pp$ collisions and simulated with PYTHIA and EPOS models. Two configurations are used for EPOS, corresponding to the mini-plasma option being respectively {\it on} (full circles) and {\it off} (open circles)~\cite{Abreu:2007kv}.}
\label{fig:EposPythia63LambdaOverKRatio}
\end{figure}

In Fig.~\ref{fig:EposPythia63LambdaOverKRatio}, we compare extrapolations at LHC energies for minimum bias $pp$ obtained with PYTHIA and the EPOS model which uses parton-based Regge theory~\cite{Drescher:2000ha} and directly includes quantum mechanical multiple scattering~\cite{Werner:2005jf;Porteboeuf:hq08}. The tune used for PYTHIA v.6.3 was already described in Ref.~\cite{Hippolyte:2006ra} and references therein. The extrapolation for EPOS at LHC energies is exactly the one reported in Ref.~\cite{Abreu:2007kv}. The presented ratios $(\rm{\Lambda}+\bar{\rm{\Lambda}})/\rm{K}$ as a function of $p_{\rm{T}}$ differ slightly (less than a 5\% effect on the overall $p_{\rm{T}}$ range) because one corresponds to neutral kaons (PYTHIA) whereas the other is for charged kaons (EPOS). The possible occurrence of mini-plasma cores leading to collective effects already in $pp$ is illustrated by the ``mini-plasma on" option: it would give a very significant baryon production with respect to meson ones for $p_{\rm{T}}$ between 2 and 9~GeV/$c$.
%
% SECTION 6: Summary
%
\section{Summary}
\label{sum_bohq08}
We presented some detector elements and their characteristics relevant for bulk matter physics at the LHC energies.
To anticipate hadro-production in $pp$ collisions at low transverse momentum, we reported several extrapolations mainly based on multiple parton interactions.
In particular, it was discussed how particle identification can help defining the structure of the underlying event in $pp$, and which hadronization mechanisms are expected to compete in A+A collisions in the intermediate transverse momentum region.
Eventually specificities of baryon and meson productions at the top LHC energies were shown based on PYTHIA and EPOS models.

\end{document}